# Rethinking the Role of Homogeneous Surface Models in High-Entropy Alloy Catalyst Screening

Jin-Soo Kim[1‡], Taegyeong Kim[1‡], Youngtak Kim[1], Sathya Sheela Subramanian[1], Geun Ho Gu[1*]

[1]Department of Energy Engineering, Korea Institute of Energy Technology, Naju 58330, Republic of Korea

## Abstract

High-entropy alloy (HEA) catalysts are commonly modeled as homogeneously mixed surfaces, although thermodynamic driving forces can produce surface segregation and chemical ordering. Here, we test the validity and practical limits of this approximation by comparing random homogeneous slabs with surfaces annealed using cluster-expansion Monte Carlo (CEMC) against three published high-throughput composition-activity datasets for the oxygen reduction and hydrogen evolution reactions. Monte Carlo-based uncertainty quantification simulations propagated uncertainties in catalyst composition and predicted binding energies through 10,000 trials. CEMC generally improved agreement with experiment, but the advantage is not robust, and neither model reliably reproduces experimental composition-activity map. Structural analysis revealed surface segregation, rather than short-range ordering, is the primary origin of the difference between the models. Nevertheless, the element-conditioned binding-energy ranges of the two models largely overlapped, supporting the use of homogeneous slabs for rapid screening of elemental combinations.

*Corresponding authors: ggu@kentech.ac.kr

‡These authors contributed equally.

# Introduction

High-entropy alloys (HEAs) have emerged as an effective materials platform for catalyst discovery owing to their compositional flexibility and the resulting diverse binding sites.[1–4] HEAs incorporate multiple elements within a single crystalline phase, providing access to a near-continuous spectrum of adsorption energies, enabling systematic exploration of catalytic trends across a broad compositional space.[5–12] The energetic diversity is expected to increase the likelihood of achieving adsorption strength near optimal values, consistent with the Sabatier principle,[13] and has motivated extensive theoretical[6–9,14–17], and high-throughput experimental efforts[10,15,18–22] to accelerate catalyst design.

Despite these advances, predictive modeling of HEA catalysts has largely relied on the assumption of homogeneous atomic mixing[6–9,11,12,14,23–25], effectively treating the surface as a direct extension of the bulk composition. While this approximation simplifies the representation of multicomponent systems, it neglects the thermodynamic driving forces that promote surface segregation and short-range ordering. Both computational studies[26–30] and emerging atomic-scale characterization[31–34] have shown that multi-metallic systems frequently deviate from random configurations, particularly at surfaces where differences in elemental surface energies reshape local environments. Recent computational work[35] has begun to incorporate segregation-driven surface structures into catalytic modeling, but their predictive performance has not yet been systematically evaluated through comparison with composition-dependent experimental activity data. Consequently, the validity of the homogeneous surface approximation for catalyst screening remains unresolved.

These observations raise two questions for computational catalyst discovery: First, are catalytic predictions governed by the nominal bulk composition of a high-entropy alloy, or by the thermodynamically selected surface state that emerges under realistic conditions? Second, if a homogeneously mixed surface model fails to predict composition-dependent experimental catalytic activity, does this mean that the homogeneous surface model is entirely useless for computational catalyst screening? Resolving these questions is critical, as most computational screening frameworks assume homogeneously mixed surfaces despite the expectation of surface segregation and short-range ordering in multicomponent alloys. More broadly, this issue extends beyond high-entropy alloys to multicomponent catalysts, where surface composition and local atomic environments are set by thermodynamic equilibration rather than nominal composition.

Cluster-expansion Monte Carlo (CEMC) can provide thermodynamically realistic surface configurations by accounting for surface segregation and chemical ordering.[35] However, experimentally synthesized HEA surfaces may retain kinetically trapped configurations rather than reaching complete thermodynamic equilibrium, and the extent of atomic rearrangement during synthesis and cooling remains uncertain.[34,36] Moreover, activity predictions based on CEMC-annealed structures still depend on approximate binding energy (BE) prediction models and simplified Sabatier relationships for activity prediction. A direct comparison with experiment is therefore required to establish both the capabilities and the limitations of these computational approaches.

Here, we compare CEMC-annealed surfaces with homogeneous mixing slabs using published high-throughput composition–activity datasets for the oxygen reduction reaction (ORR)[19,37] and hydrogen evolution reactions (HER)[15]. Because both the measured catalyst compositions and predicted BE are subject to uncertainty relative to their true values, we evaluated the models using MC-based uncertainty quantification (UQ), which is the systematic propagation of uncertainties in model inputs into distributions of predicted outputs and model-performance metrics.[38] In the present study, UQ is used to determine whether the apparent difference between the two surface models remains meaningful after uncertainties in catalyst composition and BE prediction are considered.

We find that the homogeneously mixed slab model generally fails to reproduce the experimentally observed variation in catalytic activity with composition. Incorporating thermodynamically driven segregation through CEMC improves agreement with experiment, indicating that surface redistribution is important for describing composition-dependent catalytic behavior. Nevertheless, CEMC-annealed structures combined with BE prediction models and Sabatier-based activity prediction models do not fully reproduce the experimental activity landscapes. At present, neither approach provides a sufficiently reliable substitute for experiments when predicting activity changes caused by variations in atomic ratio within a fixed multielement system. Despite this limitation, homogeneous slabs may retain practical value at an earlier stage of catalyst discovery: they can be used to screen whether a given combination of elements can provide binding environments near the catalytic optimum. This distinction defines a useful boundary for homogeneous models—they may assist in selecting constituent elements, but they should not be expected to reliably optimize their atomic ratios.

# Results

**Assessment of Annealed and Homogeneous Surface Models with Uncertainty Quantification**

The atomic-scale structure of the HEA surfaces investigated here—whether random or chemically ordered—remains uncertain. Among the three experimental studies considered, studies 1[19] and 2[37] synthesized the HEA catalysts by magnetron sputtering without intentional heating or cooling, suggesting relatively slow cooling that may allow partial annealing and structural equilibration, whereas study 3[15] employed rapid cooling from 2023 K to room temperature within 5 s. Previous experimental and kinetic Monte Carlo (KMC) studies provide contrasting indications regarding the structures retained after cooling. Kim et al.[34] used scanning tunneling microscopy to observe local chemical order at the surface of a CoCrFeMnNi HEA that was water-cooled from 3500 K to room temperature. In contrast, the KMC simulations of Du et al.[36] indicated that bulk crystal of CrCoNi medium-entropy alloys can retain their high-temperature structures when cooled rapidly from 1500 to 600 K over 10 s. Whether atomic rearrangement occurs on comparable timescales remains unclear especially for surface; the open surface may facilitate faster atomic diffusion and thereby accelerate ordering.

Accordingly, we evaluated whether CEMC-annealed slabs or randomly mixed homogeneous slabs more accurately reproduce the experimentally measured catalytic activities, while leaving the determination of the exact kinetically trapped structures outside the scope of this study. For the CEMC model, the annealing temperature was selected based on the best agreement according to Kendall's $\tau_b$ describing the rank order, mean squared error (MSE) assessing the activity in the absolute scale, and continuous ranked probability score (CRPS) measuring the similarity between the distributions. Since both the catalyst compositions and the predicted binding energies (BE) are subject to uncertainty, we performed uncertainty quantification Monte Carlo (UQMC) simulations to statistically compare the predictive performance of the two structural models. Composition uncertainty was quantified from the discrepancy between inductively coupled plasma (ICP) measurements and the nominal synthesis inputs, which may arise from evaporation or sublimation during synthesis. BE uncertainty in the linear regression model trained on DFT data was estimated from the discrepancy between DFT calculations and experimental measurements. The composition uncertainty, or composition error, defined as nominal minus ICP ratio, was approximated by a Gaussian distribution, $\mathcal{N}(0.000,\ 0.0469^2)$. The BE uncertainty, or BE error, defined as theory minus experiment and expressed in eV, was modeled separately for OH and H as $\mathcal{N}(0.042,\ 0.26^2)$ and $\mathcal{N}(-0.009,\ 0.189^2)$, respectively. Further details are provided in the Methods section.

**Figure 1** compares the predictive performance of CEMC and homogeneous slabs for the experimental ORR activity of study 1[19], study 2[37], and HER activity of study 3[15]. We evaluated the agreement with metrics—$\tau_b$, MSE, and CRPS—between the experimental and predicted activities as 3D histograms across 10,000 UQMC trials. We marked the most probable trial for both models with a red star in **Figure 1**. In the case of study 2[37], we independently performed UQMC for each micro-library (ML) 1~ML6. For the study 3[15], we examined FCC (111), (110), (100) facets with varying active sites. The facet with highest probability may be considered as the active site. **Figure 1b** shows only ML4 among the six MLs, whereas **1c** shows only the FCC (100) hollow site. Results for the remaining MLs and facet/site are presented in **Figure S6-S7**. The BE and composition errors of the most probable trials selected from the UQMC for each of the three studies are summarized in **Table S1~S2**.

Across the three studies, the CEMC model generally shows high probability at higher $\tau_b$, lower MSE and CRPS values than the homogeneous model, indicating better agreement with the experiment. For study 1, the UQMC results in **Figure 1a** show that the ($\tau_b$, MSE, CRPS) coordinates of the most probable trial, marked by a star, were (0.32, 0.19, 0.35) for CEMC and (0.03, 0.34, 0.50) for the homogeneous. For study 2 (**Figure 1b**), the ($\tau_b$, MSE, CRPS) coordinates were (0.064, 0.03, 0.004) for CEMC and (0.46, 0.05, 0.004) for the homogeneous. **Figure S6** shows that every MLs except ML5 demonstrates slightly better reproduction of the experiment with CEMC model. For these sets, CEMC shows concentrated probability distribution in the 3D histogram, whereas homogeneous shows a diffuse broad probability over larger regions indicating low distinguishability. For the FCC (100) hollow site in study 3 (**Figure 1c**), the ($\tau_b$, MSE, CRPS) coordinates were (0.22, 0.06, 0.02) for CEMC and (-0.37, 0.11, 0.01) for the homogeneous. As shown in **Figure S7**, none of the other sites showed significantly better agreement with the experimental results for CEMC than for the homogeneous model. Unlike studies 1 and 2, both the CEMC and homogeneous models exhibited diffuse posterior in the 3D histograms for study 3.

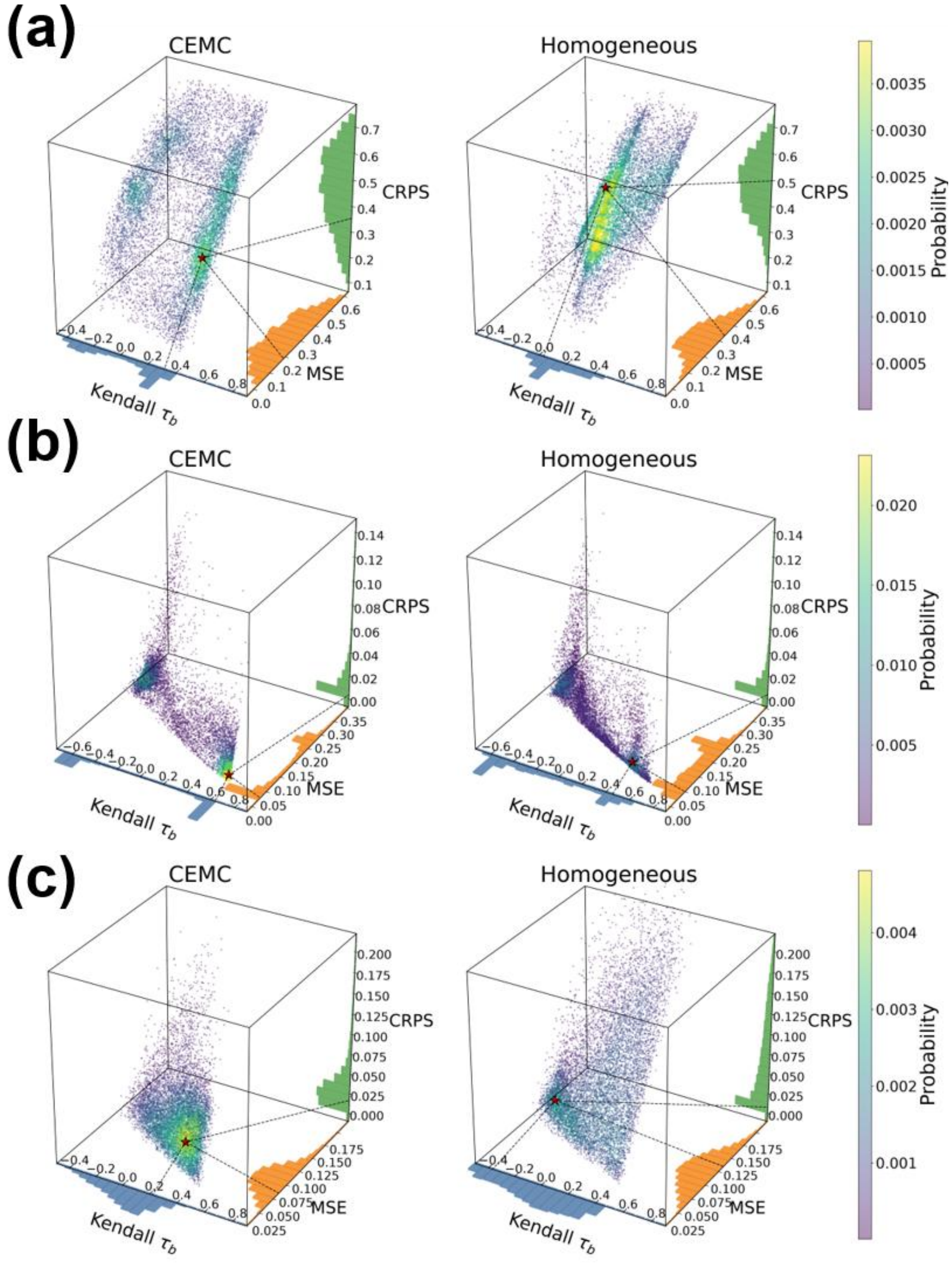

**Fig. 1** Uncertainty quantification Monte Carlo (10,000 trials) results for CEMC-annealed and homogeneous slab. Kendall $\tau_b$, mean squared error (MSE), continuous ranked probability score (CRPS) 3D histogram probability map for (a) Study 1[19] (b) Study 2[37] and (c) Study 3[15]. The trial corresponding to the most probable bin in the 3D histogram is marked with a red star indicating its coordinate by dashed lines.

**Table 1** summarizes, for each metric, the probability that CEMC agrees more closely with the experimental activity than the homogeneous model across the UQMC trials for the three studies. The positive Δ values indicate that CEMC agrees more closely with the experimental data than the homogeneous model. In particular, the random-selection baseline probability that the Δ values for each metric are positive is 1/2, or 0.5, and the average values across the three studies are 0.60, 0.57, and 0.51 for $\tau_b$, MSE, and CRPS, respectively. Thus, CEMC improves largely $\tau_b$, followed by MSE and CRPS, while CRPS shows almost no improvement over the baseline. For ML2, ML3, and ML5 in study 2, the probabilities of an improvement in CRPS were 0.36, 0.26, and 0.38, respectively – well below 0.5. The random-selection baseline probability all three metrics are positive is 1/8, or 0.125, and the minimum value of this probability was 0.17 in all three studies (The average across the three studies is 0.29). The 95% confidence intervals for the Δ metrics indicate that a substantial fraction of the trials yielded negative Δ values. The broad 95% confidence intervals, which frequently extend below zero, indicate that the advantage of CEMC is not robust across all UQMC trials.

**Table 1**. Probability that CEMC agrees more closely with the experiments than the homogeneous model in studies 1[19], 2[37], and 3[15]. Each metric is defined as follows: $\Delta\tau_b = \tau_{\mathrm{CEMC}} - \tau_{\mathrm{Homo}}$, $\Delta\mathrm{MSE} = \mathrm{MSE}_{\mathrm{Homo}} - \mathrm{MSE}_{\mathrm{CEMC}}$, $\Delta\mathrm{CRPS} = \mathrm{CRPS}_{\mathrm{Homo}} - \mathrm{CRPS}_{\mathrm{CEMC}}$. The values in brackets indicate the 95% confidence interval (from 2.5% to 97.5% percentile) for each metric, $\Delta\tau_b$, ΔMSE, and ΔCRPS.

| | $p(\Delta\tau_b > 0)$ | $p(\Delta\mathrm{MSE} > 0)$ | $p(\Delta\mathrm{CRPS} > 0)$ | $p(\Delta\tau_b > 0, \Delta\mathrm{MSE} > 0, \Delta\mathrm{CRPS} > 0)$ |
|---|---|---|---|---|
| Study 1 | 0.46<br>[-0.66,0.39] | 0.50<br>[-0.17, 0.23] | 0.51<br>[-0.16, 0.27] | 0.32 |
| Study 2, ML1 | 0.58<br>[-0.30, 0.89] | 0.63<br>[-0.07, 0.14] | 0.64<br>[-0.04, 0.05] | 0.40 |
| Study 2, ML2 | 0.78<br>[-0.22, 0.54] | 0.55<br>[-0.07, 0.06] | 0.36<br>[-0.06, 0.04] | 0.26 |
| Study 2, ML3 | 0.67<br>[-0.45, 0.68] | 0.54<br>[-0.09, 0.08] | 0.26<br>[-0.09, 0.03] | 0.18 |
| Study 2, ML4 | 0.55<br>[-0.72, 1.01] | 0.49<br>[-0.13, 0.14] | 0.50<br>[-0.04, 0.04] | 0.22 |
| Study 2, ML5 | 0.56<br>[-0.18, 0.25] | 0.59<br>[-0.05, 0.06] | 0.46<br>[-0.05, 0.04] | 0.17 |
| Study 2, ML6 | 0.71<br>[-0.29, 0.80] | 0.64<br>[-0.06, 0.12] | 0.38<br>[-0.05, 0.03] | 0.29 |
| Study 3,<br>fcc(100) hollow | 0.57<br>[-0.36, 0.54] | 0.69<br>[-0.05, 0.08] | 0.68<br>[-0.06, 0.16] | 0.36 |
| Study 3,<br>fcc(100) bridge | 0.60<br>[-0.40, 0.47] | 0.53<br>[-0.06, 0.07] | 0.53<br>[-0.08, 0.13] | 0.28 |
| Study 3,<br>fcc(110) bridge | 0.61<br>[-0.27, 0.37] | 0.58<br>[-0.04, 0.07] | 0.54<br>[-0.06, 0.14] | 0.30 |
| Study 3,<br>fcc(111) hollow | 0.44<br>[-0.61, 0.53] | 0.45<br>[-0.08, 0.08] | 0.53<br>[-0.10, 0.13] | 0.23 |
| Study 3, fcc(111) top | 0.65 [-0.24, 0.38] | 0.60 [-0.06, 0.09] | 0.68 [-0.07, 0.17] | 0.44 |
| Average | 0.60 | 0.57 | 0.51 | 0.29 |

**Figure 2** shows the composition-activity maps obtained from the experiment and simulations for the three studies. As shown in left panel of **Figure 2a**, the sputtered combinatorial library reported by study 1[19] reveals extended regions of low activity rather than the smooth activity gradient expected from continuous compositional changes. This observation implies that catalytic behavior is governed by specific surface ensembles rather than composition-averaged chemistry. Experimental activities were min-max normalized prior to comparison with simulations to preserve relative activity trends, as scanning electrochemical cell microscopy (SECCM) used in the experiment lacks a well-defined electrochemically active surface area. The simulated composition–activity maps were generated using the most probable UQMC trial (red star in **Figure 1**) for each of the CEMC and homogeneous models.

**Figure 2a** shows that the homogeneous model substantially overestimates the activities of compositions with low experimental activity (near 0). The CEMC model also tends to overestimate the activities of compositions with intermediate experimental activity (around −0.5), although this error is less pronounced than that of the homogeneous model. **Figure 2b** shows that both CEMC and the homogeneous model overestimate the low-activity experimental region (pixels on the left). In the high-activity experimental region (pixels at the bottom), CEMC predicts high activity, whereas the homogeneous model predicts only moderate activity. **Figure 2c** shows that CEMC correctly assigns high activity to the experimentally high-activity region (upper pixels), whereas the homogeneous model predicts low activity in this region. Both models overestimate the activities of experimentally low-activity compositions. The activity maps for the remaining MLs in Study 2 and the remaining facet/site combinations in Study 3 are presented in **Figures S8–S9**. The composition-activity map generated from the best-performing trial of each model for all three studies are shown in **Figure S10**. The 3D histograms show that, for study 1 and ML1, ML2, and ML5 of study 2, the best-performing CEMC trials yield more favorable $\tau_b$ and MSE values than the corresponding homogeneous-model trials, however, with low probability, and unusually large perturbation. This trend is also evident in the activity maps shown in **Figure S10**. The activity maps therefore indicate that the agreement of both models remains unsatisfactory, although CEMC agrees slightly better with the experimental data than the homogeneous model.

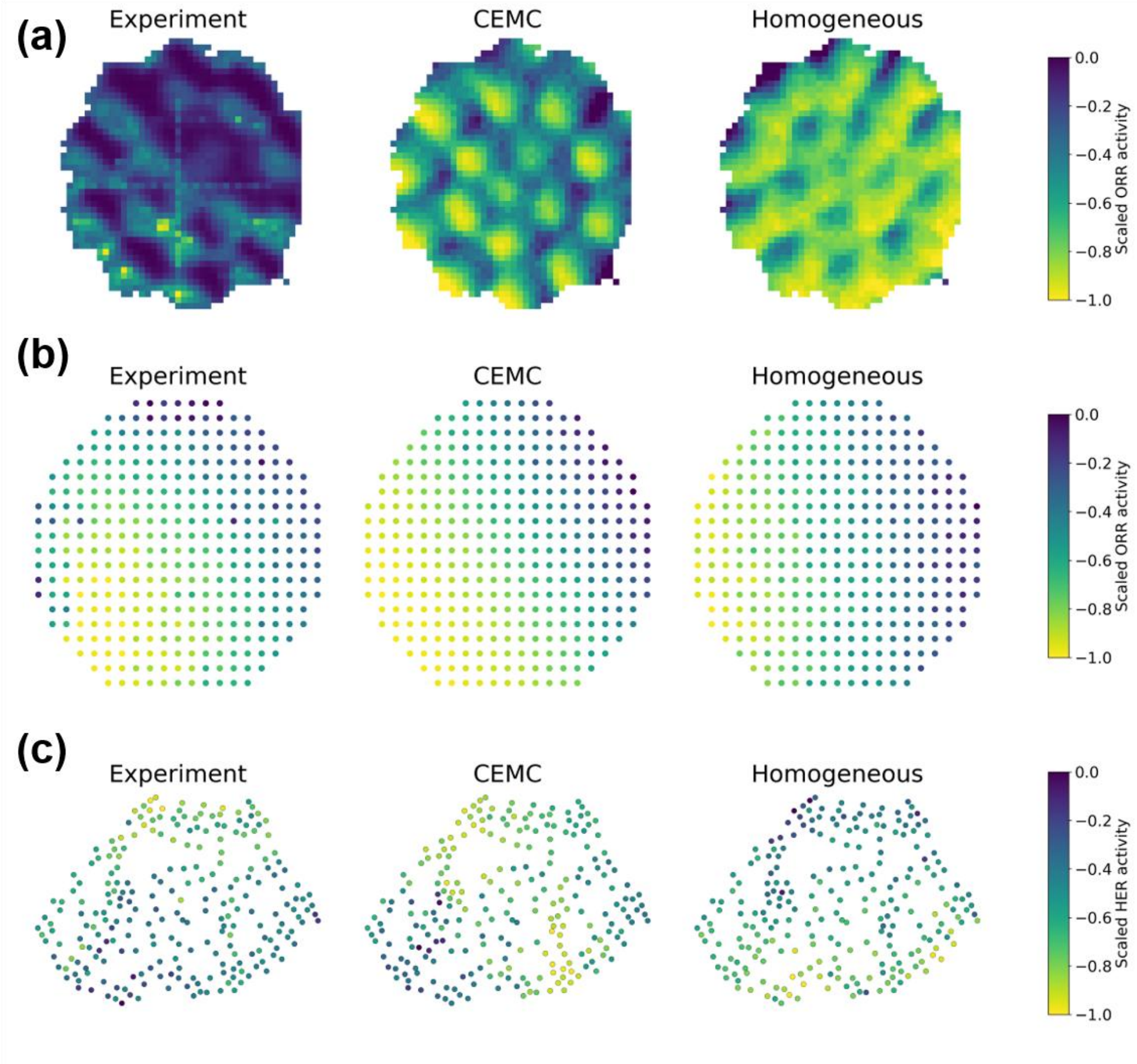

**Fig 2**. Composition-activity map obtained from experiment, CEMC, and homogeneous model with their most probable trial (marked as a red star in **Fig. 1**) for (a) Study 1, (b) Study 2, and (c) Study 3. Each pixel represents a distinct composition, and each map comprises 1400, 342, and 238 pixels for study 1, 2, and 3, respectively.

From a catalyst screening perspective, the ability to identify promising compositions is important. Thus, we present the continuous recall analysis for three studies, which were obtained using this most probable trial in each model, as shown in **Figure S11-S12**. For example, when the top 20% of samples are selected from the entire compositions based on experimental activity, recall is defined as the fraction of those samples whose predicted activities also fall within the top 20%. CEMC achieved slightly higher recall than the homogeneous model in Study 1; ML1, ML3, ML4, and ML6 of Study 2; and the FCC(100) hollow site of Study 3. In study 1 (**Figure S11f**), recall of homogeneous model approaches to the random baseline. However, in study 2 (**Figure S10**), the homogeneous model often achieves recall values above the baseline, and the difference between CEMC and the homogeneous model is small. In study 3 (**Figure S11a-e**), except for the FCC (100) hollow site, both CEMC and the homogeneous model perform similarly to the random baseline. This may suggest that the FCC (100) hollow site is the active site of the FeCoNiPdPt HEA catalyst, and the homogeneous model may fail to screen promising composition ratios at this site. The activity maps reveal persistent systematic errors in both models, particularly the overestimation of experimentally low-activity compositions. CEMC more closely reproduces some experimental trends than the homogeneous model, but substantial discrepancies remain for both approaches. We therefore next examine the structural origin of CEMC's relative advantage and the potential role of homogeneous slabs in catalyst screening.

**Surface segregation rather than local ordering reshapes the active-site landscape** Performing CEMC simulations results in higher surface fractions of Pd and Pt than those observed in homogeneous slabs in PtPdIrRhRu system, and higher Pd fraction in FeCoNiPdPt system, as shown in **Figure 3a** and **3b**. The temperature-dependent compositional evolution (**Figure 3a** and **3b**) shows that the outermost layer is dominated by Pd and Pt, indicating strong surface enrichment of elements associated with favorable segregation energetics. At elevated temperature, the surface retains partial mixing, reflecting the stabilizing contribution of configurational entropy. Upon cooling to 298 K, however, the surface undergoes pronounced chemical stratification, becoming strongly enriched in Pd and/or Pt, while the rest are largely excluded from the surface layer.

**Figure S13** shows the layer-resolved composition of the CEMC-annealed slab at 298 K for the equi-atomic PtPdIrRhRu fcc(111), FeCoNiPdPt fcc(100), FeCoNiPdPt fcc(110), and FeCoNiPdPt fcc(111) surfaces. In PtPdIrRhRu system, the second layer becomes Ru- and Rh-rich (36.5, 30.7%, respectively) with minor contribution of Pd (12.5%), Ir (10.6%), and Pt (9.7%). Third and fourth layers display Ir-, Ru-, and Rh-rich compositions, while Pd and Pt contribution is small. In the 5th layers, the Pd, Pt, and Rh fractions are close to their bulk values, whereas Ru and Ir are slightly depleted relative to the bulk composition. These trends indicate that surface segregation of Pd and Pt, and subsurface segregation of Ru and Rh after CEMC. In the FeCoNiPdPt system, the second layer of all three facets displays Fe-rich composition, followed by Pt contribution on fcc(100) and fcc(110), and Co contribution on fcc(111). In the third and fourth layers of all three facets, elements that were minor constituents of the surface and subsurface layers become dominant. Because Pd dominates the surface layer, it becomes a minor constituent in the second through fifth layers. Across all three facets, Co and Ni show a tendency to occupy third layer and deeper layers.

Such preferential enrichment induces short-range ordering within the surface layer (**Figure 3c** and **3d**). To quantify this effect, we evaluated the Warren-Cowley short-range order parameter[28]

$$\alpha_{ij} = 1 - \frac{P_{ij}}{x_j}$$

where $P_{ij}$ is the probability of finding a $j$ atom on the 1st nearest-neighbor shell of an $i$ atom on the surface, $x_j$ is the surface concentration of $j$. The resulting correlations reveal strongly pair-dependent ordering behavior. As shown in **Figure 3c**, Rh-Pt pair exhibit increasingly negative Warren-Cowley parameters with decreasing temperature, indicating preferential association and the formation of Pt-rich local environments. In contrast, Pd-Rh pair display positive values that grow with decreasing temperature, reflecting chemical separation between these elements. Notably, Pd-Pt, Pt-Pt, and Pd-Pd pairs remain close to zero across the temperature range, indicating near-random mixing and weak pairwise preference.

Fe-Pt pair show negative values, whereas Fe-Co, Co-Ni, and Fe-Ni pairs exhibit positive values (**Figure 3d**). Co-Pd, Pd-Pt, Fe-Pd, Co-Pt, and Ni-Pt pairs show values close to zero. Thus, Pd, which constitutes most of the surface composition, is randomly distributed without a clear tendency to associate with or avoid other elements. In contrast, the small amounts of Fe and Pt segregated at the surface tend to cluster together, whereas Fe, Co, and Ni show a strong tendency to avoid one another. These results demonstrate that surface ordering is not characterized by uniform clustering but by selective pairwise interactions that stabilize specific local motifs. The catalytic interface should therefore be viewed as a thermodynamically selected ensemble of chemically distinct environments rather than a statistically random mixture.

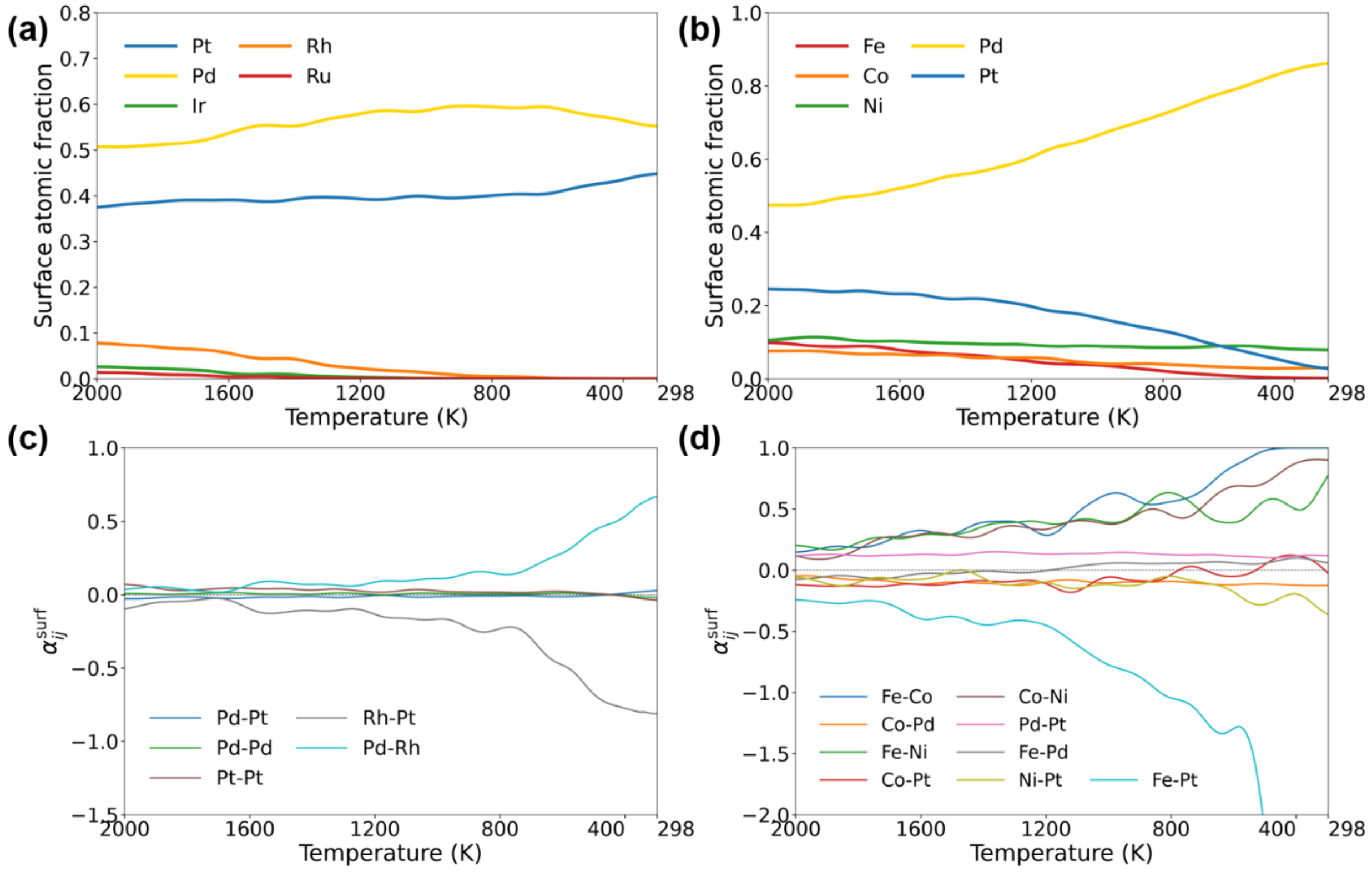

**Fig. 3** Thermodynamic surface reconstruction induces segregation and ordering. Temperature-dependent evolution of surface composition for the equal ratio surface of (a) PtPdIrRhRu fcc(111) and (b) FeCoNiPdPt fcc(100). Temperature-dependent Warren-Cowley parameters for selected pairs on (c) PtPdIrRhRu fcc(111) and (d) FeCoNiPdPt fcc(100) surfaces.

To quantify the surface redistribution effect, we define the surface compositional deviation,

$$D = \frac{\sum_i^N \left| c_i^{\mathrm{surf}} - c_i \right|}{2 - \frac{2}{N}}$$

where $c_i$ and $c_i^{surf}$ denote the system, surface compositions of element $i$, and $N$ is the number of elements, respectively. The denominator is a normalization factor to normalize the surface compositional deviation to unity. For a homogeneously mixed slab with an equi-atomic surface layer, $D$ is zero. Increasing $D$ therefore reflects stronger thermodynamic redistribution of elements toward energetically preferred surface configurations. For the equi-atomic systems annealed by CEMC at 298 K, the $D$ values were 0.75 for PtPdIrRhRu FCC(111), 0.84 for FeCoNiPdPt FCC(100), 0.94 for FeCoNiPdPt FCC(110), and 0.82 for FeCoNiPdPt FCC(111). Therefore, thermodynamic annealing redistributes the surface elements markedly different from random mixing.

This redistribution reshapes the adsorption-energy landscape (**Figure 4**). As shown in **Figure 4a**, whereas homogeneous surfaces exhibit a broad multimodal distribution, CEMC annealing produces dominant peak near 1.0 and 1.2 eV, close to the Sabatier optimum for OH adsorption and primarily associated with Pt- and Pd-containing ensembles. Consequently, the predicted activity increases, from 0.047 to 0.14. This contraction indicates that thermodynamic reconstruction selectively amplifies catalytically favorable motifs while suppressing unfavorable configurations. Randomly shuffling atoms within each CEMC layer, which removes local ordering while preserving surface segregation, produces a similar BE distribution. Thus, the differences between the CEMC and homogeneous models arise primarily from surface segregation rather than surface ordering.

As shown in **Figure 4b**, the H-BE distributions associated with the different elements have similar peak heights and positions, on homogeneous FeCoNiPdPt FCC(100) hollow site. CEMC annealing amplifies distribution associated with Pd-containing ensembles, thereby the predicted activity increases, from 0.088 to 0.19. Layer shuffled CEMC slab displays slightly different distribution with slightly increased activity of 0.038. The H-BE distributions for the remaining facet/site combinations of the FeCoNiPdPt system are presented in **Figure S14** and show trends similar to those observed for the FCC(100) hollow site. Thus, surface segregation alters the BE distributions, whereas surface ordering appears to have only a minor effect.

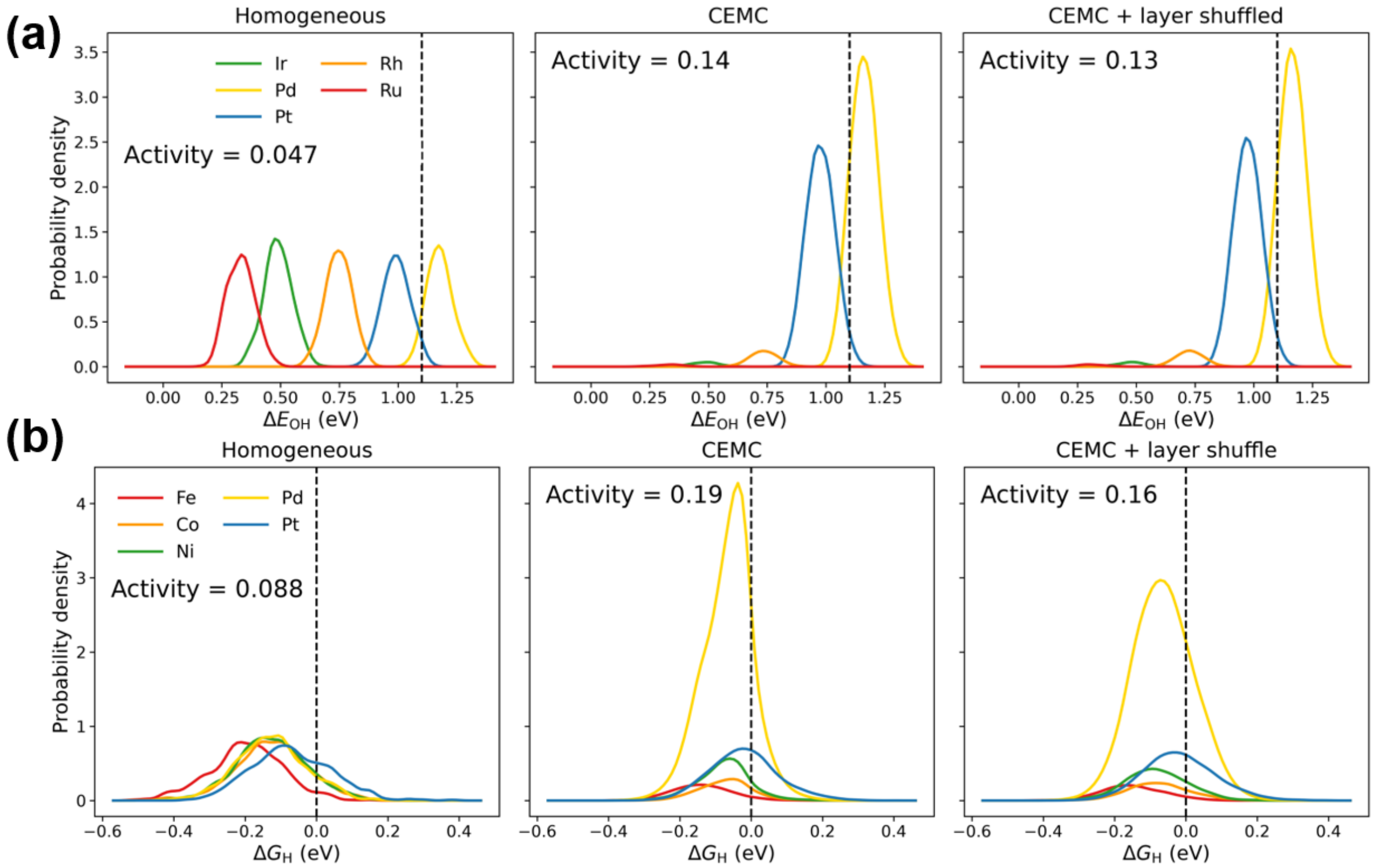

**Fig. 4** Surface-composition-weighted probability density of binding energy for equi-atomic homogeneous, CEMC-annealed, and CEMC with layer shuffled slabs of (a) PtPdIrRhRu fcc(111) and (b) FeCoNiPdPt fcc(100). Optimum binding energy is marked with dashed line.

**Is homogeneous slab model useful for screening?** Although the homogeneous slab model is less accurate than the CEMC-annealed slab model in predicting changes in catalytic activity with elemental fractions within a given compositional system, it may still be useful for selecting, for example, five elements for an HEA catalyst from a broader candidate pool, such as 20 transition metals. Selecting five elements from 20 yields 15,504 possible combinations, and screening these combinations using homogeneous slabs is substantially faster than performing MC annealing for every candidate system. The number of active sites with BEs close to the optimum does not change substantially relative to that in the homogeneous slab, although CEMC alters the elemental composition of the surface. If an active site is defined as a site with a BE within ±0.1 eV of the optimal value, the activity of a slab increases with the number of such sites. For example, for the equi-atomic PtPdIrRhRu composition, **Figure 5a** shows the element-resolved OH BE distributions for the homogeneous slab, and the corresponding activity volcano. **Figure 5b** exhibits the H-BE distributions on the equi-atomic FeCoNiPdPt fcc(100) hollow site with activity volcano. **Figure 5** indicates that only sites with BEs within 0.1 eV close to the optimal value contribute meaningfully to the activity. Thus, the activity is primarily determined by whether sites with BEs close to the optimal value are present.

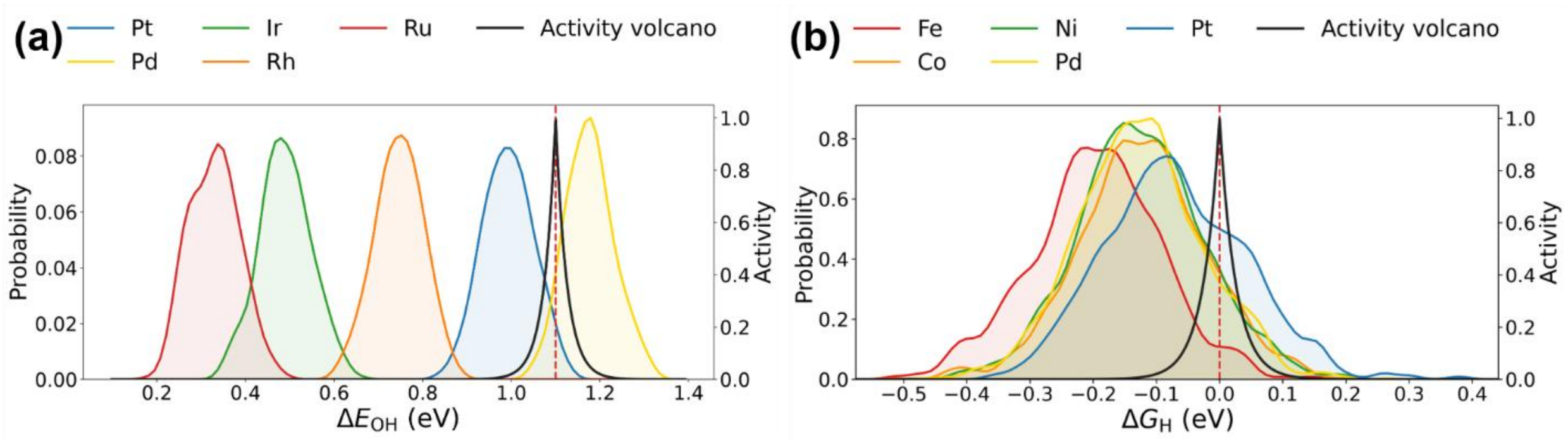

**Fig. 5** Binding energy distribution for equi-atomic homogeneously mixed (a) PtPdIrRhRu fcc(111) surface (OH adsorption on top site) (b) FeCoNiPdPt fcc(100) surface (H adsorption on hollow site). An activity volcano curve, determined using the scaled value of $e^{-\frac{|\Delta E_i - \Delta E_{opt}|}{k_B T}}$, centered at the optimal binding energy is also shown.

When the BE distributions of the CEMC slab and homogeneous slab are compared solely based on whether they contain the active site, the two distributions are nearly identical. **Figure 6** presents the unweighted, element-specific BE distributions, which represent the probability that a selected surface atom of a given element has a particular BE, irrespective of its surface fraction. For every element, more than 90% of the CEMC BE distribution overlaps with the corresponding homogeneous slab BE distribution. For both CEMC and the homogeneous model, 20 independent slabs were generated, and their BE distributions are shown in **Figure 6**. In addition, 20 slabs were independently sampled 10,000 times for each model to estimate the 95% confidence intervals for the containment percentage of each element; at least 90% for most systems as summarized in **Table S3**. Therefore, homogeneous slabs alone may be sufficient to screen whether a given elemental combination can produce active-site environments.

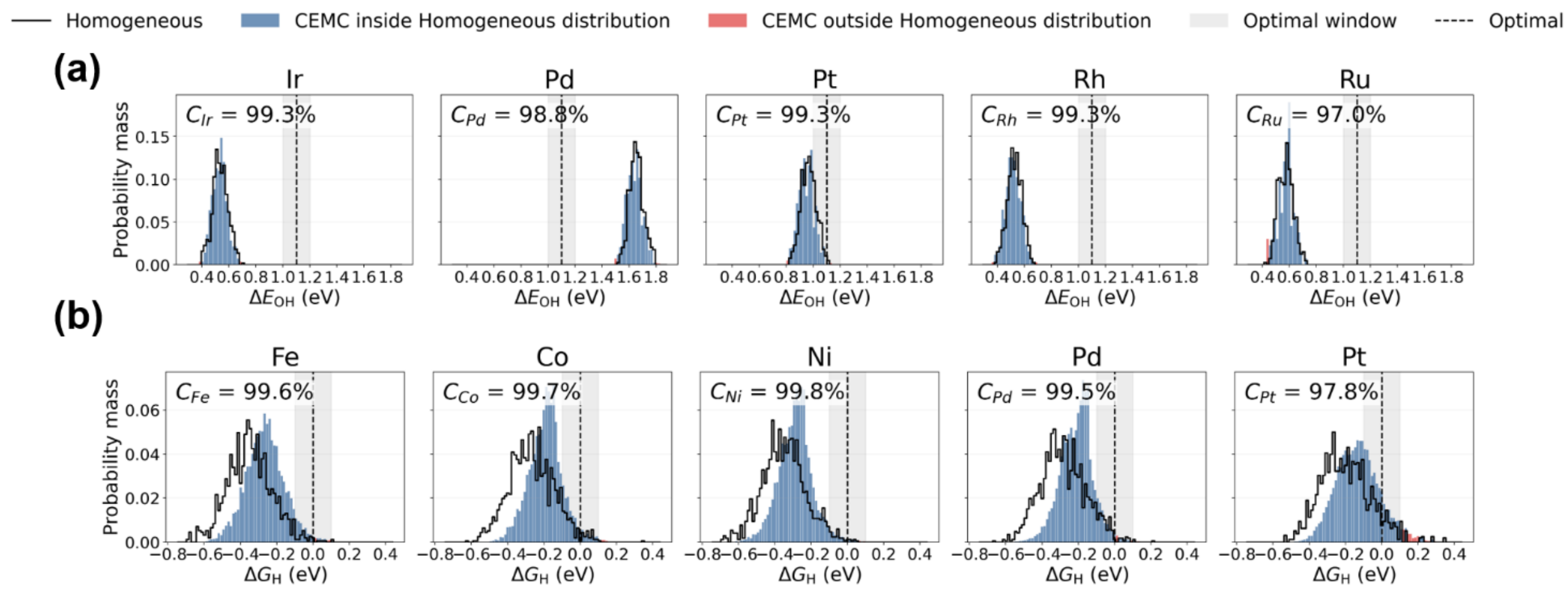

**Fig. 6** Element-conditioned probability mass of binding energy (BE) for equi-atomic homogeneous and CEMC slabs on (a) PtPdIrRhRu fcc(111) top and (b) FeCoNiPdPt fcc(100) hollow site. Blue and red bars indicate the portions of the CEMC distributions lying inside and outside the corresponding homogeneous BE distribution ranges, respectively. $C_e$ of element $e$ represents the percentage of CEMC-derived BE values for element $e$ that fall within the range sampled by the homogeneous slab.

**Figure 7** illustrates the workflow of utilizing homogeneous slab for computational catalyst screening. For example, to screen five-element combinations selected from 20 transition metals, homogeneous slabs could be constructed for each combination at the equi-atomic composition. After calculating the surface BE distributions, the number of active sites with BEs within ±0.1 eV of the optimal value could be compared across the combinations. The combinations could then be ranked by the number of active sites, providing guidance on which candidates should be prioritized for further analysis to determine whether the elements associated with these active sites will segregate to the actual surface. If these elements are confirmed to segregate to the surface, their experimental synthesis could be considered. However, within a fixed five-element system, changes in catalytic activity resulting from variations in the atomic fractions cannot currently be reliably screened using either the homogeneous or CEMC model and must therefore be evaluated experimentally. The remaining discrepancies for both models suggest that experimental activity landscapes are influenced by effects not captured in the present framework.

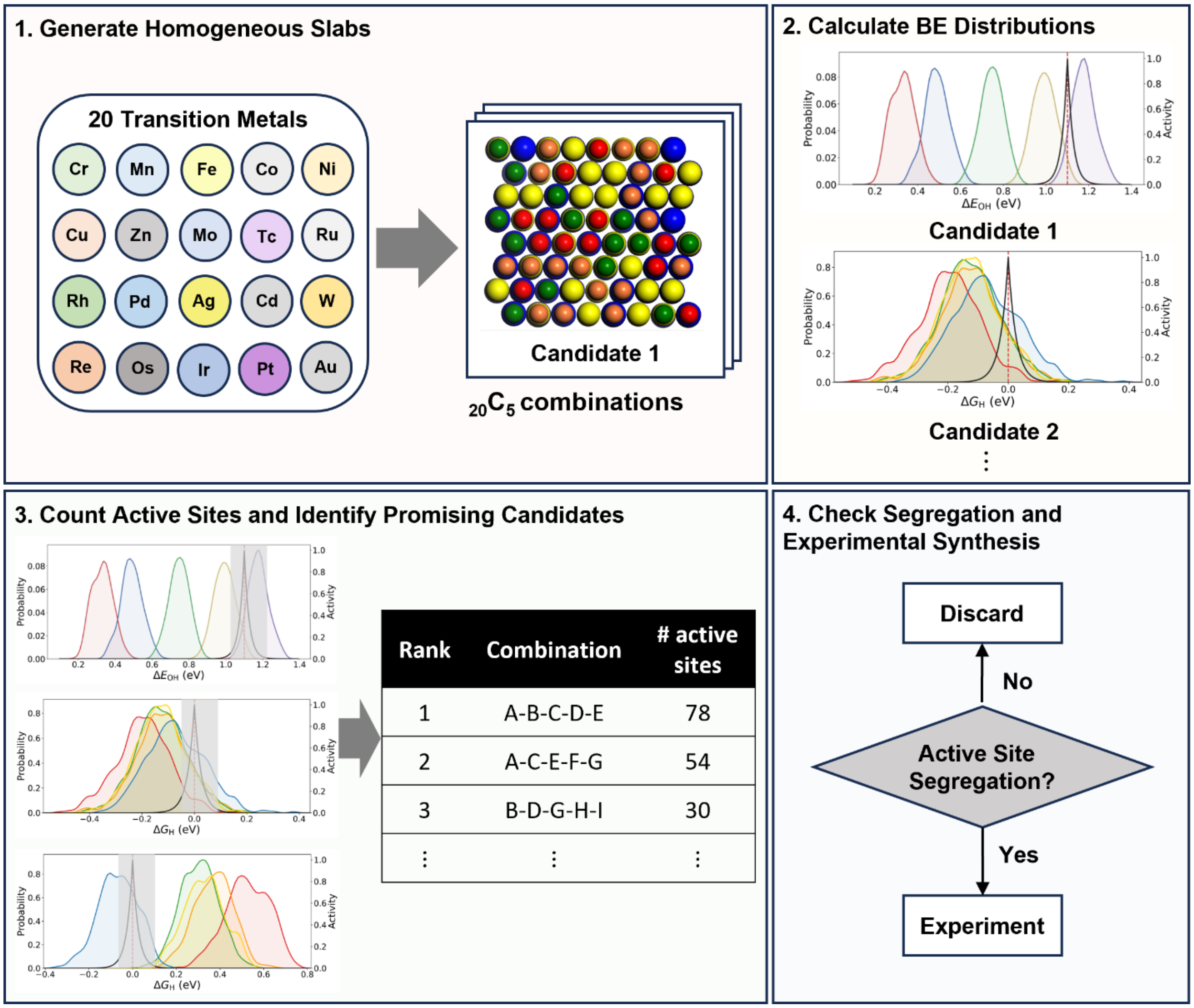

**Fig. 7** Schematic workflow of utilizing homogeneously mixed slab model for early stage of computational catalyst discovery.

**Limitations and scope of the model** Several approximations were adopted in the quantitative description of the experimental system. Catalytic activity was estimated from thermodynamic overpotentials derived from scaling relations and therefore neglects kinetic barriers and uncertainties associated with the scaling relationships. Although recent study[39] have shown that coverage-dependent lateral interactions between adsorbates can affect composition–activity trends, we assumed isolated adsorption sites to simplify the problem. According to Du et al.[36], the degree of ordering in bulk depends on the annealing duration requiring kinetics; however, the present work only accounts for the thermodynamics. In PtPdIrRhRu system, calculations were restricted to the (111) surface, whereas multiple facets may be present experimentally. The omission of kinetic barriers, coverage-dependent adsorbate interactions, annealing kinetics, and experimentally relevant facet distributions may therefore account for part of the remaining discrepancy between the predicted and experimental activity landscapes.

## Conclusions

We systematically evaluated whether thermodynamically annealed HEA surfaces reproduce composition-dependent catalytic activity more accurately than conventionally assumed homogeneous surfaces. Across the three experimental datasets,[15,19,37] CEMC was generally more likely to achieve higher $\tau$ and lower MSE and CRPS than the homogeneous model, with mean probabilities of 0.60, 0.57, and 0.51, respectively. However, residual errors in the activity maps and recall gains show that the improvement is limited; neither model yet reliably predicts composition-activity relationship. CEMC revealed pronounced Pd and Pt surface enrichment and selective short-range ordering. Layer-wise shuffling largely preserved the BE distributions, demonstrating that surface segregation, rather than short-range ordering, is the primary origin of the differences between the two models for describing activity. Nevertheless, homogeneous slabs captured most element-conditioned BE ranges and may therefore remain useful for rapid, early-stage screening of elemental combinations. While the considered experiments exposed the active sites, the screened candidate may not segregate the elemental dependent active sites, thus surface thermodynamics is a critical final step in HEA screening. Our results point that the experiments remain necessary for quantitative composition optimization. The limited absolute agreement of both models indicates that additional physical effects beyond thermodynamic surface reconstruction must be incorporated to achieve quantitatively reliable activity predictions.

## Methods

**DFT Calculation** All spin-polarized DFT calculations were implemented using the Vienna Ab initio Simulation Package (VASP)[36], with the revised Perdew-Burke-Ernzerhof (RPBE) exchange-correlation functional[37] within the generalized gradient approximation (GGA) where the core electrons are treated with the projector augmented-wavefunction (PAW) method.[38,39] Plane-wave energy cutoffs were set to 500 eV for the bulk calculations of pure metals (Ru, Rh, Pd, Pt, Fe, Co, Ni, and Ir), and 400 eV for randomly mixed HEA bulk and slab configurations. All atomic structures were generated using Atomic Simulation Environment (ASE).

For pure metal bulks calculations, the self-consistent loop convergence and force convergence criteria were set to $10^{-8}$ eV and 0.01 eV Å$^{-1}$. Brillouin zone sampling was performed using a 15 × 15 × 15 Monkhorst-Pack *k*-point grid[40] with a Gaussian Methfessel-Paxton smearing factor of 0.1.[41]

For randomly mixed HEA bulk and slab calculations, the self-consistent loop convergence and convergence criterion for force were set to $10^{-5}$ eV and 0.05 eV Å$^{-1}$, respectively. Monkhorst-Pack grids[40] of 4 × 4 × 4 and 3 × 3 × 1 were used for HEA bulk and slab calculations, respectively. A 10 Å vacuum layer was introduced along the surface normal to avoid interactions between periodic slab images.

For the calculation of H adsorption energies on Fe-Co-Ni-Pd-Pt surfaces, the initial magnetic moments were set to 4, 2, 1, 1, and 1 $\mu_B$ for Fe, Co, Ni, Pd, and Pt, respectively. A configuration was excluded from the training and test sets of the H adsorption energy prediction model if (i) the magnetic moment of any atom differed by more than 1 $\mu_B$ between the clean and H-adsorbed slabs, (ii) the absolute magnetic moment of the H atom exceeded 0.1 $\mu_B$, or (iii) the H atom was displaced by more than 0.5 Å in the xy plane relative to its initial position.

**H adsorption energy prediction model** For the Fe-Co-Ni-Pd-Pt system, H adsorption energy was estimated using the linear model proposed by Batchelor et al.[9]:

$$\Delta E_{pred} = \sum_{z}^{zones} \sum_{j}^{sites} C_{z,j} N_{z,j}$$

Here, $C_{z,j}$ and $N_{z,j}$ are the model parameters for zone $z$ and the number of each site $j$ found in each zone. The linear models were trained separately, using distinct datasets for five adsorption sites across three facets: the top and hollow sites on fcc(111), the bridge and hollow sites on fcc(100), and the long-bridge site on fcc(110). For the top and hollow sites on fcc(111), we adopted without modification the model architecture proposed by Batchelor et al.[9] for predicting OH and O adsorption energies on the Ru-Rh-Pd-Pt-Ir system. Values for the fcc(111) top/hollow prediction model parameters are shown in **Table S4-S5**. For the fcc(100) and fcc(110), the zone definition and model parameters are presented in **Figure S2**, **Table S6-S8**. The accuracy of trained linear models is presented in **Figure S3**. The root-mean-square-error (RMSE) against the DFT testing

data are 0.048, 0.067, 0.055, 0.045, and 0.114 eV for fcc(111) top/hollow, fcc(100) hollow/bridge, and fcc(110) long-bridge site, respectively.

**Formation energy prediction model** For Ru-Rh-Pd-Pt-Ir system, 658 configurations of 3 × 3 × 6 FCC (111) slabs and 1,100 configurations of $2 \times 2 \times 2$ FCC bulk structures were generated with randomly distributed Ru, Rh, Pd, Pt and Ir atoms to construct the training dataset. For Fe-Co-Ni-Pd-Pt system, 1,100 configurations of $2 \times 2 \times 2$ FCC bulk structures were generated with randomly distributed Fe, Co, Ni, Pd and Pt atoms to construct the common training dataset for FCC (111), (110), and (100) slabs. For each facet, 700, 691, and 1,019 3 × 3 × 6 FCC (111), (110), and (100) slab configurations were generated. The lattice constants of the HEA bulk and slab configurations were set to the arithmetic mean of the DFT-relaxed lattice constants of the constituent pure metals.[9]

The cluster expansion (CE) model was trained to predict the formation energies of multicomponent alloy configurations. The model takes the initial unrelaxed atomic structure as input and predicts the DFT energy of the relaxed configuration. Each input structure is mapped onto an ideal fcc parent lattice. The occupation of lattice site $i$ is represented by a discreate variable $\sigma_i$, which could take on of six states (for the Ru-Rh-Pd-Pt-Ir system),

$$\sigma_i \in \{\mathrm{Ru}, \mathrm{Rh}, \mathrm{Pd}, \mathrm{Pt}, \mathrm{Ir}, X\},$$

where $X$ denotes a vacancy. The vacancy was introduced to represent the vacuum region of the slab, thereby allowing bulk and surface configurations to be described using the same lattice-based representation.

The formation energy per metal atom of a configuration $\boldsymbol{\sigma}$ is expressed as

$$\Delta E_f^{CE}(\boldsymbol{\sigma}) = \frac{1}{N_m}\left[\sum_a J_a N_a + \sum_{s=1}^{4}\sum_{a\leq b} J_{s,ab} N_{s,ab} + \sum_g \sum_d J_{g,d} N_{g,d}\right],$$

where $N_m$ is the number of metal atoms. The four terms in brackets represent the point, pair, and triplet contributions, respectively. The coefficients $J_a$, $J_{s,ab}$, and $J_{g,d}$ are the effective cluster interactions (ECIs) fitted during model training.

The point contribution describes the dependence of the energy on the number of atoms of each metal species, where $a \in \{\mathrm{Ru, Rh, Pd, Pt, Ir}\}$. Here, $N_a$ is the number of atoms of species $a$, and $J_a$ is its fitted point coefficient. The point terms depend only on the overall composition and therefore cannot distinguish configurations having the same composition but different atomic arrangements. Such configurational differences are described by the pair and triplet terms. No separate point term was assigned to $X$; the effects of vacancies were instead represented by metal-vacancy pair and vacancy-containing triplet terms.

In the pair contribution, $N_{s,ab}$ is the number of unordered $a - b$ pairs in the $s$-th FCC neighbor shell and $J_{s,ab}$ is the corresponding fitted pair coefficient. The shell index was defined as $s = (r/r_{NN})^2$, where $r_{NN}$ is the nearest-neighbor distance. Pair interactions through the fourth neighbor shell, $s$=1,…,4 were included. The $X$-$X$ pair was excluded because clusters occupied entirely by vacancies do not contribute to the model.

In the triplet contribution, $g$ specifies the geometry of a three-site cluster and $d$ specifies its chemical decoration. Chemical decoration is the assignment of chemical species or vacancies to the three vertices of a given triplet geometry, after accounting for the symmetry of the geometry. For example, Pt-Pt-Pd, Pt-Pd-Rh, and Pt-Ru-X are different chemical decorations of a three-site cluster. The quantity $N_{g,d}$ is the number of triplets with geometry $g$ and decoration $d$, and $J_{g,d}$ is the corresponding fitted triplet coefficient. Two triplet geometries were included: (1,1,1) and (1,1,2). The three integers specify the neighbor-shell indices of the three edges of the triplet. Thus, a (1,1,1) triplet is an equilateral nearest-neighbor triangle whose three edges all belong to the first neighbor shell. A (1,1,2) triplet is an isosceles triangle with two first-neighbor edges and one second-neighbor edge. The all-vacancy triplet is excluded.

The ECIs were trained using ridge regression. For each configuration $i$, the normalized cluster counts were assembled into a descriptor vector $A_{i,\alpha} = N_{i,\alpha}/N_{m,i}$, where $N_{i,\alpha}$ is the number of occurrences of decorated cluster $\alpha$ and $N_{m,i}$ is the number of metal atoms. Before fitting, each descriptor column was divided by its root-mean-square value, $q_\alpha = \left[n_{fit}^{-1} \sum_i A^2{}_{i,\alpha}\right]^{1/2}$, giving the scaled descriptor matrix $\widetilde{A_{i,\alpha}} = A_{i,\alpha}/q_\alpha$; this places frequently and rarely occurring cluster functions on comparable numerical scales. The scaled coefficients were obtained from

$$\hat{\beta} = \mathrm{argmin}_{\beta} \left\{ \sum_i \left( y_i - \sum_\alpha \widetilde{A_{i,\alpha}} \beta_\alpha \right)^2 + \lambda \sum_\alpha \beta_\alpha^2 \right\},$$

where $y_i$ contains the DFT formation energy of configuration $i$ and $\lambda$ controls the regularization strength. Since $\beta$ multiplies the scaled descriptors, the ECIs multiplying the original cluster functions are $J_\alpha = \beta_\alpha / q_\alpha$, such that $\widetilde{\mathbf{A}}\boldsymbol{\beta} = \mathbf{AJ}$. The regularization parameter was selected by five-fold, bulk/slab-stratified cross-validation on the 90% fitting set from 31 logarithmically spaced values between $10^{-8}$ and $10^{2}$, using the mean validation MAE as the selection criterion. The selected $\lambda$ values were $10^{-8}$ for PtPdRhRuIr–(111), 46.4159 for FeCoNiPdPt–(100), 21.5443 for FeCoNiPdPt–(110), and 2.15443 for FeCoNiPdPt–(111). After holdout evaluation, the final models were refitted to all available reference configurations using the corresponding selected $\lambda$. The test results for 10 % holdout set for the four datasets are presented in **Figure S4**.

**Uncertainty Quantification Monte Carlo** Uncertainty was propagated through the composition–structure–activity workflow by Monte Carlo sampling, independently for total eight sets of compositions: 1,400 compositions of Ru-Rh-Pd-Pt-Ir ORR of the study 1[19], six materials library datasets (ML1-ML6; 342 compositions in each ML) of the study 2[37], and the study 3[15] (238 compositions of Fe-Co-Ni-Pd-Pt HER). For each dataset, 10,000 UQ trials were performed over all compositions. For Studies 1 and 2, one composition error per element was sampled from a Gaussian distribution in each trial. For Study 3, 238 composition errors per element—one for each composition—were independently sampled from the same distribution in each trial. This UQMC scheme reflects the experimental procedures: hundreds of catalysts in Studies 1 and 2 were synthesized simultaneously by magnetron sputtering, whereas the 238 catalysts in Study 3 were synthesized individually. The Gaussian distribution was derived from 12 published papers[40–51] reporting both the nominal elemental ratios used during synthesis and the corresponding compositions measured by inductively coupled plasma (ICP) analysis for materials containing the relevant constituent elements (see **Table S9**). For each reported element, the compositional deviation was calculated as

$$\Delta x = x_{\mathrm{nominal}} - x_{\mathrm{ICP}},$$

where $x_{\text{nominal}}$ and $x_{\text{ICP}}$ are the nominal and ICP-measured atomic fractions, respectively. The resulting deviations were pooled to construct the empirical error distribution shown in **Figure S5a**. This distribution was fitted with a Gaussian function, yielding a mean of 0.000347 and a standard deviation of 0.046858 in fractional-composition units. For each sampled composition, the drawn error was subtracted from the nominal fraction, consistent with the definition above. Negative fractions were clipped to zero, and all elemental fractions were subsequently renormalized so that their sum equaled unity.

The BE error distribution was derived from the various literature sources,[52–64] which contains both theoretical and experimental adsorption energies on various surfaces. For each data point, the prediction error was defined as

$$\Delta\Delta E = \Delta E_{\text{theory}} - \Delta E_{\text{experiment}}$$

The BE errors for H and OH were independently compiled from the literature to construct separate error distributions, which were subsequently fitted to Gaussian distributions. The fitted H BE error distribution had a mean of 0.000 eV and a standard deviation of 0.1893 eV, whereas the OH BE error distribution had a mean of 0.04233 eV and a standard deviation of 0.2604 eV (see Supplementary **Figure S5b** and **S5c**).

For each UQ trial, an independent $\Delta\Delta E$ was sampled for every constituent element and treated as a trial-level systematic model error shared across all compositions. For OH adsorption at a top site occupied by element $e$, $\Delta E_{\text{OH}}^{\text{UQ}} = \Delta E_{\text{OH}}^{\text{model}} - \Delta\Delta E_e$. For H adsorption at a site formed by multiple metal atoms, the correction was the arithmetic mean of the element-specific errors of the directly binding atoms (zone 1): $\Delta E_{\text{H}}^{\text{UQ}} = \Delta E_{\text{H}}^{\text{model}} - (1/N_{\text{zone1}}) \sum_{i=1}^{N_{\text{zone1}}} \Delta\Delta E_{e_i}$.

For each perturbed composition, 20 independent slabs were randomly generated and evaluated both before annealing (random-alloy reference) and after cluster-expansion Monte Carlo (CEMC). Snapshots were evaluated at 18 temperatures from 2000 to 300 K in 100 K increments. The activity of each slab was computed and the 20 activities for each composition and temperature. Within each trial, predicted and experimental activities across the all compositions were separately min–max scaled to [-1, 0], with the favorable endpoint mapped to -1. Agreement with experiment was quantified using Kendall's rank correlation coefficient ($\tau_b$), mean squared error (MSE), and continuous rank probability score (CRPS).

**Activity Agreement Metrics** For each composition $i = 1, \dots, N$, the experimental and computational activities, both scaled to the range [-1, 0], were denoted by $x_i$ and $y_i$, respectively. Rank agreement was quantified using $\tau_b$, which accounts for tied values:

$$\tau_b = \frac{C - D}{\sqrt{(C + D + T_x)(C + D + T_y)}},$$

where $C$ and $D$ are the numbers of concordant and discordant composition pairs, respectively, and $T_x$ and $T_y$ denote the number of pairs tied only in the experimental and computational activities. Numerical prediction errors were evaluated using the MSE,

$$\text{MSE} = \frac{1}{N}\sum_{i=1}^{N}(y_i - x_i)^2.$$

To calculate CRPS, experimental and computational activities were divided into the same activity bins, and the normalized bin counts were cumulatively summed to obtain their empirical cumulative distributions, $F_{\text{exp}}$ and $F_{\text{calc}}$. The CRPS was then calculated as

$$\text{CRPS} = \sum_{k=1}^{K}\left[F_{calc}(z_k) - F_{exp}(z_k)\right]^2 \Delta z,$$

where $z_k$ is the k-th activity bin and $\Delta z$ is the bin width.

**Cluster expansion Monte Carlo** The Ru-Rh-Pd-Pt-Ir surface was modeled as an 10 × 10 × 10 FCC(111) slab. The Fe-Co-Ni-Pd-Pt surface was modeled as 10 × 10 × 10 FCC(111), (110), and (100) slab. Atomic configurations were optimized using Metropolis Monte Carlo annealing. Each Monte Carlo step consists of exchanging the atomic species of two randomly selected lattice sites. The acceptance criterion for a trial move is given by the Boltzmann distribution:

$$e^{\frac{E(s_{new}) - E(s)}{k_B T}} \geq random(0,1) \tag{1}$$

where $s$ represents the current configuration, $s_{new}$ denotes the configuration after the atomic exchange, $k_B$ is the Boltzmann constant, $T$ is the simulation temperature, and $random(0,1)$ indicates a random number between 0 and 1. The energy is estimated using the pretrained cluster expansion model for the relevant system and facets.

Configurational entropy is implicitly captured through the Monte Carlo sampling process, as exchange moves are sampled across all possible atomic configurations, and their acceptance probabilities reflect the multiplicity of accessible states. A warm-up simulation was first performed at the initial temperature (2023 K) for 10,000 steps, followed by gradual cooling to 298K over 100,000 steps.

**Catalytic Activity Assessment** For each slab configuration, the OH or H adsorption energy was estimated using the linear adsorption model proposed by Batchelor *et al.*[9] and above-mentioned method. The adsorption energies were then converted into catalytic activity using a transition-state-theory-inspired expression based on the Sabatier principle:

$$A = \ln\left(\frac{k_\mathrm{B}T}{hN}\sum_{i=1}^{N}\exp\left(-\frac{\left|\Delta E_\mathrm{opt} - \Delta E_i\right|}{k_\mathrm{B}T}\right)\right) \tag{2}$$

where $N$ is the total number of active sites on the surface, $h$ is the Planck constant, $\Delta E_\mathrm{opt}$ is the optimal OH or H adsorption energy, and $\Delta E_i$ is the OH or H adsorption energy of slab at site *i*. The optimal OH adsorption energy, $\Delta E_\mathrm{OH,opt}$, of 1.1 eV was obtained from previous studies by Stephens *et al.*[42] The optimal H adsorption energy, $\Delta E_\mathrm{H,opt}$ was derived from the HER volcano relationship proposed by Norskov et al.[65], according to which HER activity is maximized when the hydrogen adsorption free energy, $\Delta G_H$, approaches to 0 eV. We determined a site-specific $\Delta E_\mathrm{H,opt}$ for each facet and adsorption geometry. For this purpose, H was adsorbed at the relevant sites of four-layer 2×2 FCC Pt slabs, and the vibrational frequencies of the adsorbed H atom were calculated using DFT. The Gibbs free energy correction, $\Delta G_{corr} = \Delta G_H - \Delta E_H$, was subsequently evaluated at 298K using VASPKIT.[66] Since the thermodynamic optimum satisfies $\Delta G_\mathrm{H,opt} = 0$, the corresponding optimal DFT adsorption energy was obtained as $\Delta E_\mathrm{H,opt} = -\Delta G_{corr}$. This procedure yielded optimal adsorption energies of -0.11 and -0.14 eV for the top and fcc-hollow sites on fcc(111), respectively, -0.08 eV for the long-bridge site on fcc(110), and -0.07 and -0.16 eV for the hollow and bridge sites on fcc(100), respectively. The overall workflow for for activity evaluation on annealed surfaces is illustrated in Error! Reference source not found..

# Acknowledgement

This work was supported by the Korea Institute of Energy Technology Evaluation and Planning(KETEP) and the Ministry of Climate, Energy & Environment(MCEE) of the Republic of Korea (No. 20224000000320). This work was supported by the National Research Foundation of Korea, the Ministry of Science and ICT under award numbers RS-2024-00429941. This research was financially supported by the Ministry of Trade, Industry and Energy (MOTIE) and Korea Institute for Advancement of Technology(KIAT) through the International Cooperative R&D program No.P0027799. We acknowledge the Korea Institute of Science and Technology Information (KISTI) for the computational resources provided for this research.